\documentclass[runningheads]{llncs}

\usepackage{eccv}

\usepackage{graphicx}
\usepackage{amsmath}
\usepackage{amssymb}
\usepackage{booktabs}
\usepackage{animate}
\usepackage{enumitem}
\usepackage{multirow}
\usepackage{diagbox}
\usepackage{color}

\usepackage{epigraph}
\usepackage{amsmath}
\usepackage{amssymb}
\usepackage{float}
\usepackage{diagbox}
\usepackage{enumitem}
\usepackage{ulem}
\usepackage{tabularx}
\usepackage{microtype}
\usepackage{soul}
\usepackage{marvosym}
\usepackage{tikz}
\usepackage{makecell}
\usepackage{colortbl}

\usepackage{eccvabbrv}

\usepackage{wrapfig}

\usepackage[accsupp]{axessibility}  % Improves PDF readability for those with disabilities.

\usepackage[pagebackref, breaklinks, colorlinks, citecolor = eccvblue]{hyperref}

\usepackage{orcidlink}

\newcommand{\reffig}[1]{{Fig.~\ref{#1}}}
\newcommand{\reftab}[1]{{Tab.~\ref{#1}}}

\newcommand{\refeq}[1]{{Eq.~\ref{#1}}}

\begin{document}\sloppy

% ---------------------------------------------------------------
% TODO REVIEW: Replace with your title
\title{Beat-It: Beat-Synchronized Multi-Condition 3D Dance Generation}

% TODO REVIEW: If the paper title is too long for the running head, you can set
% an abbreviated paper title here. If not, comment out.
\titlerunning{Beat-It: Beat-Synchronized Multi-Condition 3D Dance Generation}

% % TODO FINAL: Replace with your author list.
% Include the authors' OCRID for the camera-ready version, if at all possible.
\author{Zikai Huang\inst{1}\orcidlink{0009-0005-4526-440X} \and
Xuemiao Xu\inst{1,3,4}$^{(\textrm{\Letter})}$\orcidlink{0000-0002-8006-3663}\and
Cheng Xu\inst{2}$^{(\textrm{\Letter})}$\orcidlink{0000-0002-4281-6214}\and
Huaidong Zhang\inst{1}\orcidlink{0000-0001-7662-9831} \and\\
Chenxi Zheng\inst{1}\orcidlink{0009-0006-0344-2439} \and
Jing Qin\inst{2}\orcidlink{0000-0002-7059-0929} \and
Shengfeng He\inst{5}\orcidlink{0000-0002-3802-4644}
}

% % TODO FINAL: Replace with an abbreviated list of authors.
\authorrunning{Z. Huang et al.}
% % First names are abbreviated in the running head.
% % If there are more than two authors, 'et al.' is used.

% % TODO FINAL: Replace with your institution list.
\institute{
South China University of Technology, China \\
\email{xuemx@scut.edu.cn}
\and
The Hong Kong Polytechnic University, Hong Kong SAR, China \\
\email{cschengxu@gmail.com}
\and
Guangdong Engineering Center for Large Model and GenAI Technology \\
\and
Guangdong Provincial Key Lab of Computational Intelligence and Cyberspace Information\\
\and
Singapore Management University, Singapore\\
\url{https://zikaihuangscut.github.io/Beat-It/}
}

\maketitle

\begin{figure}[h]
    \centering
    \centering
    \includegraphics[width=1\textwidth]{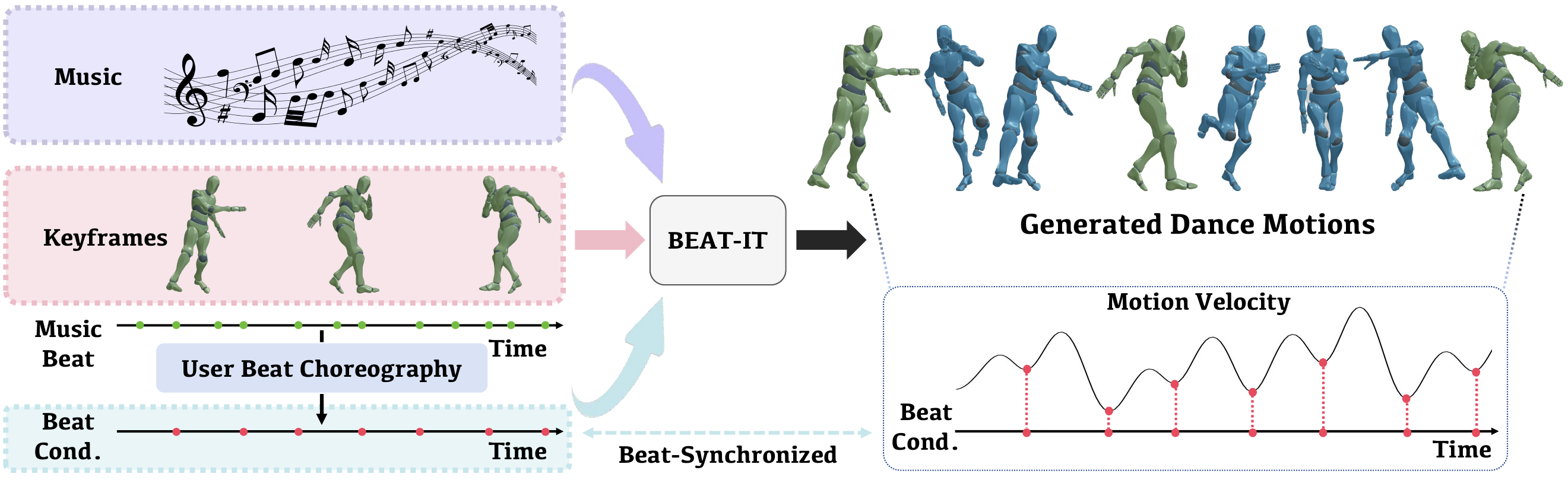}
    \caption{We introduce \textbf{\textit{Beat-It}}, a novel method for generating 3D dance motions with beat alignment and motion controllability. Our approach explicitly injects beat awareness and seamlessly integrates multiple conditions to guide the generation process, leading to beat-synchronized, key pose-guided dance generation.}
    \label{teaser}
\end{figure}

\begin{abstract}
Dance, as an art form, fundamentally hinges on the precise synchronization with musical beats. However, achieving aesthetically pleasing dance sequences from music is challenging, with existing methods often falling short in controllability and beat alignment. To address these shortcomings, this paper introduces Beat-It, a novel framework for beat-specific, key pose-guided dance generation. Unlike prior approaches, Beat-It uniquely integrates explicit beat awareness and key pose guidance, effectively resolving two main issues: the misalignment of generated dance motions with musical beats, and the inability to map key poses to specific beats, critical for practical choreography. Our approach disentangles beat conditions from music using a nearest beat distance representation and employs a hierarchical multi-condition fusion mechanism. This mechanism seamlessly integrates key poses, beats, and music features, mitigating condition conflicts and offering rich, multi-conditioned guidance for dance generation. Additionally, a specially designed beat alignment loss ensures the generated dance movements remain in sync with the designated beats. Extensive experiments confirm Beat-It's superiority over existing state-of-the-art methods in terms of beat alignment and motion controllability.

\end{abstract} 
\section{Introduction}
\label{sec:intro}
Dance, a time-honored art form, resonates across various ages and cultures due to its artistic and aesthetic significance, playing a vital role in the cultural and entertainment sectors. Traditional dance choreography creation, which demands an intricate balance of aesthetic movements, emotional expression, and precise synchronization with musical beats, often presents itself as a costly, laborious, and time-intensive endeavor, even for seasoned artists.

Recent advances in deep learning, particularly in generative models, have catalyzed efforts to automate the choreography process. Initial approaches predominantly focused on single-condition dance generation, learning a direct mapping from music to dance motions~\cite{GrooveNet, melody, revolution, danceformer, aist++, tsmt, bailando, transflower}. The advent of diffusion models, renowned for their exceptional content generation capabilities, has marked a paradigm shift in this domain. For instance, Tseng et al.~\cite{edge} introduced a transformer-based diffusion model that sets a new benchmark in music-to-dance conversion, showcasing the potential of these models in elevating the quality of generated dance sequences.

Notwithstanding the remarkable progress, achieving precise synchronization between dance motions and musical beats, coupled with flexible motion controllability in dance generation, remains a formidable challenge. 
In real-world choreography, the goal is often to assign specific key poses to particular musical beats and then fill in the connecting movements to form the complete dance sequence. 
This process requires not only the creation of designated key poses but also their harmonious synchronization with the musical beats. While few methods have explored key pose conditions~\cite{keyframe, edge}, none of them has considered explicit beat synchronization and controllability, rendering inferior dance motion generation.
Furthermore, previous methods typically merge keyframe features with music features through simple concatenation or temporal/spatial blending. This approach, however, is flawed due to the sparse nature of keyframes. Consequently, these methods disproportionately emphasize dense music features at the expense of key pose conditions, leading to a notable misalignment between the generated dance motions and the specified keyframes.

To combat the above issues, we aim to completely disentangle beats from input music and simultaneously incorporate explicit beat and key pose guidance to enable beat-controllable, key pose-guided dance generation. We introduce Beat-It, a diffusion-based, multi-condition framework designed for this purpose~(see \reffig{teaser}). Our approach begins by formulating beat condition as a nearest beat distance representation, which is a straightforward yet effective method to serve as beat guidance within our framework. To effectively harness and balance the multiple conditions involved, we propose a hierarchical multi-condition fusion mechanism. Initially, we integrate sparse key pose condition with dense beat and music conditions using a tailored beat-aware dilated cross-attention strategy. We further blend refined beat and music features to generate comprehensive multi-condition features. This strategy not only enhances beat and motion controllability in dance generation but also maintains the realism and coherence of the produced dance motions. Additionally, leveraging our beat representation, we introduce a novel beat alignment loss, explicitly ensuring synchronization between dance motions and the given beat condition, thereby significantly enhancing the overall quality of generation. Extensive experiments confirm that our method outperforms current state-of-the-art approaches in terms of beat alignment and motion controllability.
Beyond the advantages above, our framework also supports arbitrary beat designation and flexible frame assignments of key poses.

In summary, our main contributions are fourfold:
\begin{itemize}
  \item We introduce a multi-condition dance generation framework that achieves beat synchronization and enhanced motion controllability. To the best of our knowledge, our method makes the first attempt at beat-controllable key pose-guided dance generation.
  \item We present a hierarchical multi-condition fusion mechanism to effectively suppress the conflicts and fully exploit the complementary information among different conditions.
  \item We delve into the property of beats and formulate it as a nearest beat distance representation. A beat alignment loss is further tailored to offer explicit supervision signals to the generated dance motions, largely elevating the synchronization between the generated motions and the given beat conditions.
  \item Extensive experiments demonstrate our method performs favorably against the state-of-the-art approaches, especially in motion controllability and motion-beat alignment.
\end{itemize}

\section{Related Work}
\subsection{Human Motion Generation}
Human motion generation aims to generate realistic human motion automatically. 
Previous works can be generally categorized into three genres according to the input condition: text-conditioned~\cite{flame, t2mgpt, oohmg, ude, mofusion}, audio-conditioned, and scene-conditioned~\cite{humanise, imos, scenediffuser, circle}. Our work belongs to audio-conditioned human motion generation. But different from the primary audio-conditioned human motion generation,  music-to-dance generation requires a comprehensive consideration of aesthetic movements, emotional expression, and precise synchronization with musical beats, making it extremely challenging to render convincing results. Recently, researchers have widely explored the application of diffusion models in human motion generation~\cite{flame, ude, mofusion, scenediffuser, diffgesture, lda, gesturediffuclip}, which shows unprecedented performance than traditional generative models. In this paper, we get a deep insight into the beat synchronization problem and investigate a multi-condition diffusion-based scheme for controllable dance generation.

\subsection{Dance Generation}
With the emergence of deep learning~\cite{xu2022fully, xu2023pose, xie2024d3still}, dance generation has attracted lots of research interest in recent years. Previous methods have explored various frameworks with different types of backbones to achieve single-condition dance generation, including CNN~\cite{cnndance}, RNN~\cite{weakly, GrooveNet, melody, revolution}, GCN~\cite{selfdance, learn2dance}, VAE~\cite{neverstop, dancemeld}, GAN~\cite{d2m, deepdance, mnet}, and Transformer~\cite{danceformer, aist++, styletransfer, tsmt, bailando, bailando++, transflower}.
FACT~\cite{aist++} proposes a full-attention cross-modal transformer model to capture the correlations between music and dance. 
Danceformer~\cite{danceformer} presents a two-stage deterministic framework for audio-driven single-condition dance generation. It first generates key poses from the input music and then performs interpolation between the key poses. However, it falls short of arbitrary beat choreography and flexible key pose controllability due to its single-condition nature.
Bailando~\cite{bailando} utilizes separate VQ-VAEs on upper/lower half bodies and a motion GPT to map the music and seed poses to dance sequences. 
Several recent attempts have investigated the diffusion models for dance generation~\cite{edge, diffdance, bidirectionaldiffusion}, which significantly pushes the boundaries of music-to-dance performance. 
Compared to applying only a single input condition, multi-condition dance generation incorporates additional conditions beyond music into the generation process for better controllability, such as dance style labels~\cite{lda}, text~\cite{tm2d}, and keyframes~\cite{edge, keyframe}. For example, LDA~\cite{lda} introduced dance style labels as auxiliary conditions. However, it is limited to generating pre-defined styles and lacks flexibility. TM2D~\cite{tm2d} adopts a VQ-VAE based model to support dual-modal music and text-driven dance motion generation. 
Yang et al.~\cite{keyframe} employ normalizing flows to model dance motion probability distributions and utilize time embedding to achieve keyframe control.
EDGE~\cite{edge} adopts a diffusion-based framework and enables key pose control via temporal/spatial blending during inference. Although the existing methods have investigated additional conditions for better dance generation control, they typically neglect explicit guidance and supervision to align the generated motion dances with specified beats. Consequently, these methods struggle to produce both beat-specific and key pose-guided dance sequences. 
In contrast, we propose to explicitly disentangle the beat condition from music and inject flexible beat and key pose controllability into the entire generation process, significantly enhancing the beat alignment while supporting flexible frame assignments of key poses.

\subsection{Multi Condition Diffusion Generation}
Multi-condition diffusion generative models seek to inject multi-condition guidance into the generation process for versatile controllability~\cite{glide, dalle, cogview, makeascene, imagen, piti, myspot, beyondtext, artforms, dragnoise, dreamanime}. Benefiting from large-scale training data, Stable Diffusion~\cite{sd} constructs a powerful foundation for image generation. Building upon this, ControlNet~\cite{controlnet} and T2I-Adapter~\cite{t2iadapter} efficiently integrate trainable parameters into existing models, enabling the inclusion of spatially localized input conditions in a pre-trained text-to-image diffusion model. Uni-ControlNet~\cite{unicontrolnet} proposes a framework equipped with two additional control modules for multi-condition control. Qin et al.~\cite{unicontrol} present a unified model capable of handling various visual conditions, resulting in a more streamlined scheme. Codi~\cite{codi} acquire multi-modal synthesis skills through the mapping of diverse modalities into a unified space, training on lists of different multi-modal generation tasks. 
The aforementioned works typically controllable generation using pre-trained models under diverse conditions. Our work, conversely, focuses on the effective and comprehensive fusion of multiple conditions by presenting a hierarchical multi-condition fusion mechanism aimed at better integrating sparse keyframe control conditions with dense music and beat control conditions, resulting in more controllable dance generation.

\section{Method}
\subsection{Problem Definition}
Dance generation aims to create realistic dance motion sequences $\mathcal{X} = \{\mathbf{x}^1, \mathbf{x}^2, \mathbf{x}^3,\\ \cdots, \mathbf{x}^L\}$ from a given piece of music $\mathcal{C}$ with a duration of $L$. Here, $\mathbf{x}^i \in \mathbb{R}^D$ represents a human pose at the $i$-th frame, which is denoted as a $D$-dimentional vector.
In this paper, we endeavor to disentangle beat condition from the input music and simultaneously incorporate explicit beat and key pose guidance to achieve beat-synchronized, key pose-guided dance generation. 
Given a music condition $\mathcal{C}$, a sparse set of keyframes $\mathcal{X}^{ref} = \mathcal{X} \odot \mathbf{M}$, where $\mathbf{M} \in {\{0,1\}}^L$ is a temporal binary mask that specifies the assigned locations of the keyframes, and a beat condition $\mathcal{B}$, our goal is to generate a beat-specific, and key pose-guided dance motion sequence $\hat{\mathcal{X}} = \{\hat{\mathbf{x}}^1, \hat{\mathbf{x}}^2, \hat{\mathbf{x}}^3, \cdots, \hat{\mathbf{x}}^L\}$ that conforms to the music $\mathcal{C}$, consistent with reference frames $\mathcal{X}^{ref}$ at positions specified by $\mathbf{M}$, and synchronized with the beat condition $\mathcal{B}$.

\subsection{Preliminaries}
In this work, we adopt diffusion model~\cite{ddpm, glide} as the backbone of our framework. Diffusion models establish a consistent Markovian forward process that incrementally introduces noise into clean sample data $\mathbf{x}_0^{1:L} \in q(\mathbf{x}_0)$, and a corresponding reverse process that progressively eliminates noise from noisy samples. For brevity, we use $\mathbf{x}_t$ to represent the entire sequence. During the forward process, a pre-defined noise variance schedule $\beta_t$ is employed to regulate the noise increments at each step. The forward process can be formulated as follows:
\begin{equation} \label{diffusion_forward}
    q(\mathbf{x}_{1:T}|\mathbf{x}_0) = \prod_{t=1}^{T}{q(\mathbf{x}_t|\mathbf{x}_{t-1})},
    q(\mathbf{x}_t|\mathbf{x}_{t-1}) = \mathcal{N}(\mathbf{x}_t;\sqrt{1- \beta_t}\mathbf{x}_{t-1}, \beta_t\mathbf{I}).
\end{equation}
After $T$ steps, the sample data progressively transforms into a noise distribution $q(\mathbf{x}_T)$, which is usually a standard Gaussian distribution $\mathcal{N}(\mathbf{0}, \mathbf{I})$. In the reverse process, the noise is gradually removed from the noisy sample $\mathbf{x}_T$ to produce the clean sample $\mathbf{x}_0$. In our task, multi-conditional dance generation aims to model the distribution $p(\mathbf{x}_0|\mathbf{C})$ with a set of conditions $\mathbf{C}$. Following Ho et al.~\cite{ddpm}, we directly predict the clean sample $\mathbf{x}_0$ from the noise distribution $q(\mathbf{x}_T)$ with the following objective:
\begin{equation} \label{L_simple}
    \mathcal{L}_{simple} = \mathbb{E}_{\mathbf{x}_0\sim q(\mathbf{x}|\mathbf{C}), t\sim[1, T]}[\lVert {\mathbf{x}_0 - G(\mathbf{x}_t, t, \mathbf{C})} \rVert _2^2].
\end{equation}
A spatial or temporal constraint $\mathbf{x}^c$ with positions specified by a binary mask $m$ can be directly incorporated into the denoising process without additional training. This can be achieved by simply performing the following operation at every timestep:
\begin{equation} \label{diffusion_constraint}
    \mathbf{x}_t = \mathbf{x}_t^c \odot m + \mathbf{x}_t \odot (1 - m),
\end{equation}
where $\odot$ denotes the Hadamard product. By doing so, regions under constraint can be replaced with forward-diffused samples of the constraint, while the remaining regions are left unchanged. However, it is worth noting that this simplistic approach tends to render inferior results when given sparse constraints.

\subsection{Overview}
\begin{figure*}[t]
    \centering
    \includegraphics[width=\linewidth,scale=1.00]{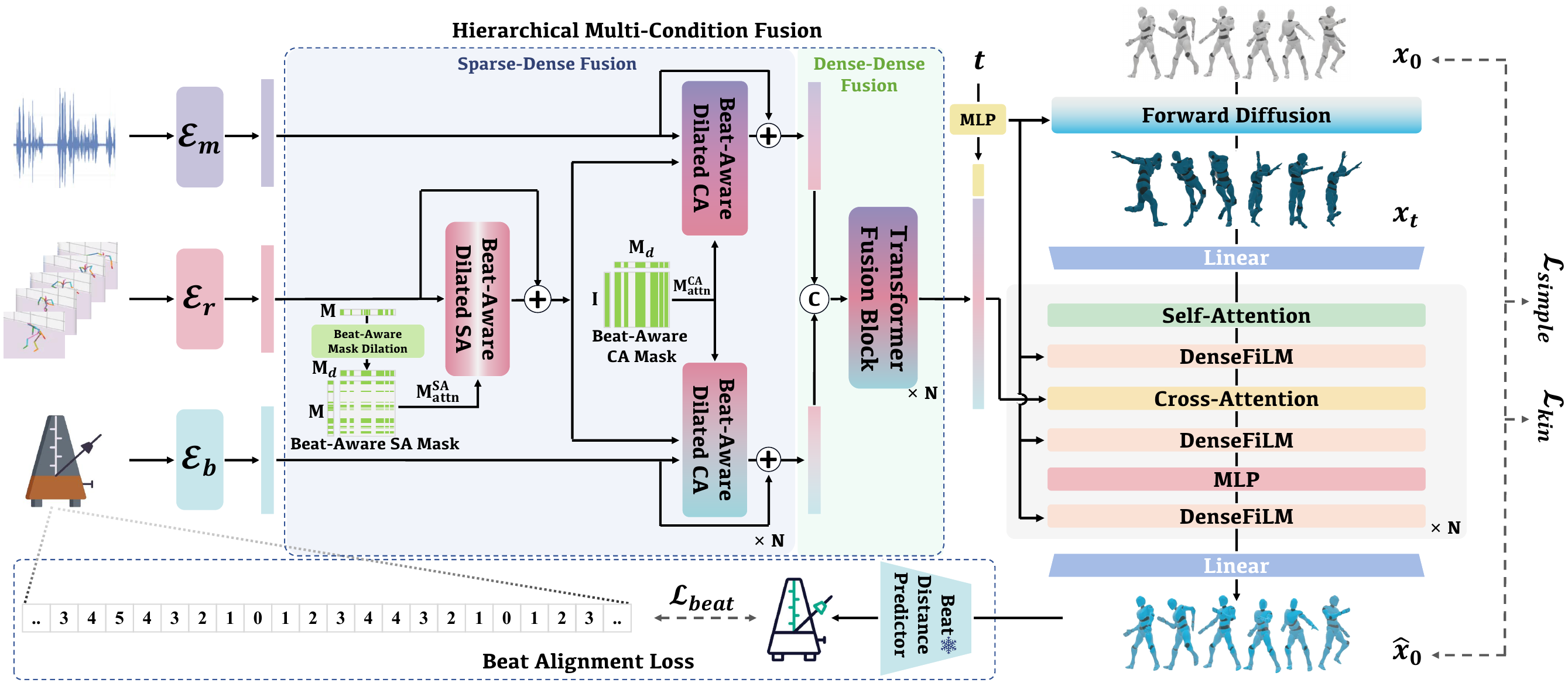}
    \caption{
        \textbf{Overview of our proposed method, \textit{Beat-It.}} 
        We generate a beat-synchronized dance sequence utilizing music, keyframes, and beat conditions. Conditional embeddings are derived and subsequently fused in a two-stage process: initially integrating sparse keyframe condition with other dense conditions, followed by the fusion of these dense conditions. The final fused condition is then processed by the conditional diffusion module. To ensure precise beat control, a beat alignment loss is employed to explicitly supervise the generated motions at the beat level.
        } 
    \label{overview}	
  \end{figure*}

The pipeline of our proposed model is illustrated in \reffig{overview}. To begin with, we feed the music condition $\mathcal{C}$, keyframe condition $\mathcal{X}^{ref}$, and beat condition $\mathcal{B}$ into three encoders respectively: the music encoder $\mathcal{E}_m$, keyframe encoder $\mathcal{E}_r$, and beat encoder $\mathcal{E}_b$. This process yields the corresponding music embedding $\mathbf{e}_m$, keyframe embedding $\mathbf{e}_r$, and beat embedding $\mathbf{e}_b$, where $\mathbf{e}_m, \mathbf{e}_r, \mathbf{e}_b \in \mathbb{R}^{L\times d}$. Then, we forward these condition embeddings to a hierarchical multi-condition fusion module to produce comprehensive multi-condition features $\mathbf{c}_f$, which are finally sent to the conditional diffusion denoising module to render the denoised dance motion sequence $\hat{\mathcal{X}}$.

\subsection{Beat Representation}
As mentioned in previous studies~\cite{aist++, bailando, edge}, there exists a strong consistency between musical beats and dance motion beats. 
In practical dance choreography, the arrangement of motion beats is relatively flexible and does not require a strict point-to-point alignment with musical beats.
Therefore, we introduce an independent beat condition disentangled from the music, to allow for flexible motion beat controllability in dance generation.
For an input music sequence of duration $L$, each frame can be categorized as either a beat or non-beat frame. 
A straightforward way to represent the beat condition is using a binary mask, where 1 and 0 denote the beat and non-beat frames respectively. However, such a simple binary presentation exhibits high sparsity as the beat frames only account for a small portion of the whole music sequence. This makes it less informative and, at times, perceived as noise or disregarded by the model. 
As an alternative, we propose to represent the beat condition as a vector $b$, with each entry $b^i \in \mathbb{N}$ denoting the distance between the current frame and the nearest beat frame. This representation not only mitigates the sparsity caused by binary formulations but also provides the model with local temporal context. This significantly facilitates the precise acquisition of rhythmic characteristics within the choreography and injects more effective beat controllability into the learning process.
The beat embedding $\mathbf{e}_b \in \mathbb{R}^{L\times d}$ is obtained using a separate embedding layer, followed by the transformer-based encoder $\mathcal{E}_b$. Please refer to the supplementary materials for detailed representations of music and dance.

\subsection{Hierarchical Multi-Condition Fusion}    
Although the keyframe constraint offers precise pose specifications for specific frames, it is still too sparse which significantly increases the learning difficulty of the model. Naively incorporating this condition with others introduces a vast of padding, leading to excessive noises that may impair the effectiveness of keyframe constraints. By contrast, music condition and beat condition are much more denser, with valid values for each frame. To harmoniously combine these three conditions, we introduce a novel multi-condition hierarchical fusion mechanism. 

\begin{figure}[t]
    \centering	
    \includegraphics[width=1\linewidth]{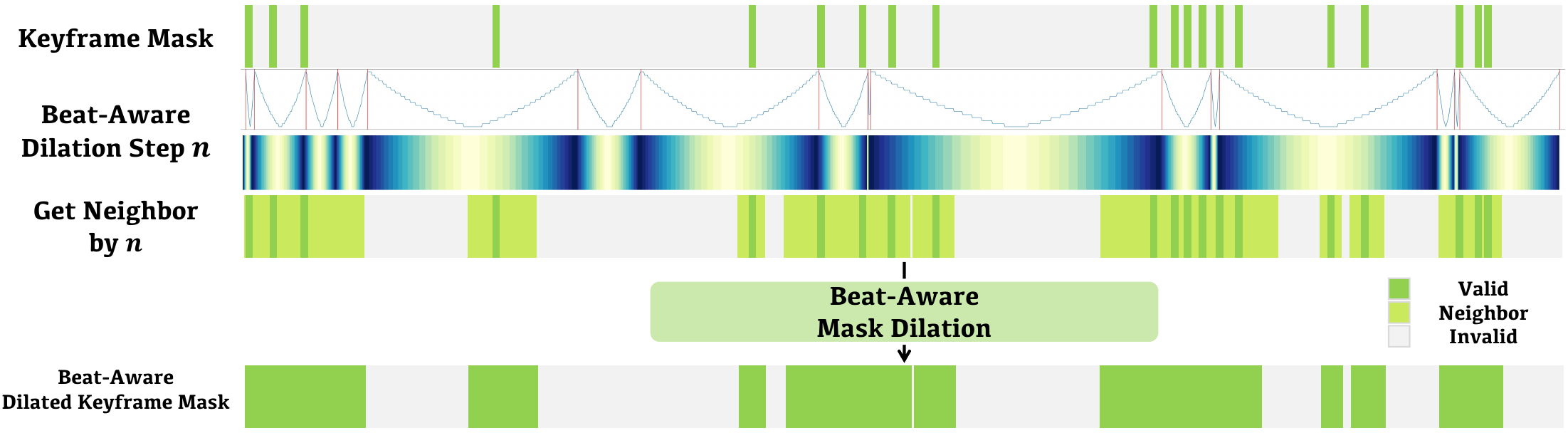}
    \caption{
        \textbf{Illustration of the beat-aware mask dilation scheme.} 
        The first row visualizes the keyframe mask, with deep green indicating valid control constraints and gray indicating invalid ones. The second row presents the dilation step curve with red lines marking beat frames. The third row is a heatmap of the dilation step. The fourth row shows the neighborhood range of keyframes, with light green indicating the expanded valid region from beat-aware mask dilation. The final row displays the beat-aware dilated keyframe mask.
    }
    \label{beat_aware_mask_dilation}	
\end{figure}
Specifically, in the first stage, we inject sparse keyframe condition into other dense conditions through a beat-aware dilation attention strategy, which shares a similar spirit of DiffKFC~\cite{diffkfc}. 
We observe that dance movements on motion-beat frames are often more iconic in the whole sequence, distinguished by their higher emphasis and significance. As these movements typically serve as guiding elements, influencing the choreography and style of the entire dance, they exert a more substantial influence on the surrounding frames. To utilize this property, we design a beat-aware mask dilation scheme. Unlike DiffKFC which uses identical dilation steps for all keyframes, treating them uniformly, we opt for different dilation weights based on the distance between the keyframe and the nearest designated beat frame.
The dilation step $n$ is calculated as $n = \lceil {s \cdot e^{-2\frac{b^i}{d^i}}} \rceil$,
where $b^i$ and $d^i$ are the ground truth beat distance at frame $i$ and the distance between frame $i$'s adjacent motion-beat frames, respectively. Base dilation step $s$ is a hyper-parameter controlling overall dilation speed. The closer the keyframe is to the nearest beat frame, the larger the dilation step. This strategy enables the model to gradually propagate the keyframe condition to the surrounding frames with beat awareness. 
The beat-aware mask dilation scheme is illustrated in \reffig{beat_aware_mask_dilation}. 
Given the keyframe binary mask $\mathbf{M}$ and dilation step $n$, the beat-aware dilated mask $\mathbf{M}^d$ is obtained as follows:
\begin{equation} \label{dilation}
    \mathbf{M}_d[i] =
    \begin{cases}
    \max_j \textbf{M}[i-j],  j\in\{-n, n+1, \dots, n-1, n\} &  \text{if $\textbf{M}_i=1$,} \\
    \textbf{M}_i,  & \text{otherwise.}
    \end{cases}
\end{equation}
The beat-aware dilated attention mask is then calculated as $\mathbf{M}_{attn} = \mathbf{M} \mathbf{M}^\intercal_d$.
This beat-aware mask dilation scheme enables beat-aware keyframe condition propagation, thereby amplifying the synergy among conditions characterized by varying degrees of sparsity, thus significantly boosting the generation quality.

In the second phase, we concatenate two refined dense conditions, namely keyframe-fused music embedding and keyframe-fused beat embedding. These concatenated conditions are fed into a transformer condition fusion module, yielding the final fused features that can be subsequently sent to the conditional diffusion denoising block.

\subsection{Beat Alignment Loss}
For controllable dance generation, it is crucial to ensure precise beat alignment between the generated dance motions and the given beat conditions. Current methods typically ignore the explicit constraints on the beats of the generated dance sequences. As a consequence, these approaches lack necessary beat controllability and struggle to guarantee beat synchronization between the generated dance motions and musical beats, which severely deteriorates the overall generation quality.
To conquer this issue, building upon our beat representation, we present a beat alignment loss to provide explicit supervision that enforces the beats of the generated motions to precisely align with the given beat conditions. To this end, we first pre-train a beat distance estimator, which accepts a motion sequence as input and predicts the distance from the current frame to the nearest motion beat frame for each frame in the sequence. During the training process, the pre-trained beat distance estimator is responsible for providing supervision signals on the beats of the generated motions. The beat alignment loss $\mathcal{L}_{beat}$ can be formulated as follows:
\begin{equation} \label{L_beat}
    \mathcal{L}_{beat} = \sum_{i=1}^{L}{w_s^i \cdot w_b^i\cdot \mathbf{MSE}(b^i, \hat{b}^i)},
\end{equation}
where $w_s^i = \frac{1}{1 + e^{a\cdot (c - \lVert{b^i-\hat{b}^i}\rVert/b^i)}}$, and $w_b^i = e^{-\frac{2b^i}{d^i}}$. Here, $\hat{b}^i$ represents the predicted beat distance at frame $i$. $w_s^i$ denotes an adaptive weight, which is derived from~\cite{shrinkage_loss}. It aims to accentuate supervision in regions where beat alignment is imprecise. This is achieved by suppressing the loss magnitude in well-aligned beat regions. This strategy enables the model to focus more attention on challenging instances. The hyper-parameters $a$ and $c$ are used for controlling the degree of penalization and the threshold for beat alignment accuracy, respectively. $\mathbf{MSE}(\cdot, \cdot)$ denotes the mean squared error. 
$w_b^i$ is an adaptive weight that enhances the supervision of frames closely aligned with the beats while diminishing the supervision of frames distantly related to the beats. Benefiting from the beat alignment loss, our model can yield dance motions with superior beat alignment and controllability, largely elevating the quality of the generated results.

\subsection{Other Losses}	
Apart from the proposed beat alignment loss $\mathcal{L}_{beat}$, we also employ the basic diffusion loss $\mathcal{L}_{simple}$ (\refeq{L_simple}) and several additional auxiliary losses to govern the training. 

\textbf{Kinematic Loss.}
We adopt the same auxiliary losses as~\cite{edge, mdm}. The joint positions loss $\mathcal{L}_{joint}$ (\refeq{L_joint}) and velocities loss $\mathcal{L}_{vel}$ (\refeq{L_vel}) are used to improve the overall physical plausibility of the generated dance motions.
\begin{equation} \label{L_joint}
    \mathcal{L}_{joint} = \frac{1}{L}\sum_{i=1}^{L}{ \lVert{FK(\mathbf{x}^i_0) - FK(\hat{\mathbf{x}}^i_0)}\rVert_2^2},
\end{equation}
\begin{equation} \label{L_vel}
    \mathcal{L}_{vel} = \frac{1}{L}\sum_{i=1}^{L}{ 
    {\lVert{{\mathbf{x}^i_0}{^\prime} - {\hat{\mathbf{x}}^i_0}{^\prime}} \rVert_2^2} + 
    {\lVert{{FK(\mathbf{x}^i_0)}{^\prime} - {FK(\hat{\mathbf{x}}^i_0)}{^\prime}}\rVert_2^2}.
    }
\end{equation}
The contact consistency loss $\mathcal{L}_{contact}$ (\refeq{L_contact}) proposed in EDGE~\cite{edge} is employed to alleviating the foot-slilding.
\begin{equation} \label{L_contact}
    \mathcal{L}_{contact} = \frac{1}{L}\sum_{i=1}^{L}{ \lVert{{FK_{foot}(\hat{\mathbf{x}}^i_0)}{^\prime} \cdot \hat{g}^i}\rVert_2^2},
\end{equation}
where $i$ denotes the index of the frame, and $FK(\cdot)$ is the forward kinematics function that converts the generated 6-DOF rotation representation into 3D key points in Cartesian space. $\hat{g}^i$ denotes the predicted binary foot contact label. In addition, we employ acceleration loss $\mathcal{L}_{acc}$ to further improve the quality of generated dance motions, which is suggested in previous work~\cite{bailando}, preventing jitters in the generated dance motions.
\begin{equation} \label{L_acc}
    \mathcal{L}_{acc} = \frac{1}{L}\sum_{i=1}^{L}{ 
    {\lVert{{\mathbf{x}^i_0}{^{\prime\prime}} - {\hat{\mathbf{x}}^i_0}{^{\prime\prime}}} \rVert_2^2} + 
    {\lVert{{FK(\mathbf{x}^i_0)}{^{\prime\prime}} - FK(\hat{\mathbf{x}}^i_0){^{\prime\prime}}}\rVert_2^2}.
    }
\end{equation}
The overall kinematic loss is formulated as follows:
\begin{equation} \label{L_kin}
    \mathcal{L}_{kin} = \lambda_{joint}\mathcal{L}_{joint} + \lambda_{vel}\mathcal{L}_{vel} + \lambda_{contact}\mathcal{L}_{contact} + \lambda_{acc}\mathcal{L}_{acc},
\end{equation} 
where $\lambda_{joint}$, $\lambda_{vel}$, $\lambda_{contact}$, and $\lambda_{acc}$ are hyper-parameters controlling the weights of the corresponding losses. We set $\lambda_{joint}=1$, $\lambda_{vel}=2.5$, $\lambda_{contact}=10$, and $\lambda_{acc}=0.1$ empirically.
The overall loss function is formulated as follows:
\begin{equation} \label{L_total}
    \mathcal{L} = \mathcal{L}_{simple} + \lambda_{kin}\mathcal{L}_{kin} + \lambda_{beat}\mathcal{L}_{beat},
\end{equation}
where $\lambda_{kin}$ and $\lambda_{beat}$ are hyper-parameters controlling the weights of the corresponding losses. $\lambda_{kin}$ and $\lambda_{beat}$ are set as 1 and 0.5, respectively.

\section{Experiments}

\textbf{Dataset.}
Following previous works~\cite{bailando, edge}, we choose the most widely used dataset, AIST++~\cite{aist++} for both training and testing. This dataset comprises 1,408 dance sequences encompassing 10 distinct street dance genres. Each dance genre incorporates 6 musical pieces with different BPMs. 85\% of these sequences are categorized as basic choreographies, where performers execute the same choreography across all 6 musical pieces, each featuring varying tempos to synchronize with distinct musical beats. With its high choreographic and beat diversity, this dataset is ideal for our controllable dance generation task.

\textbf{Implementation Details.}
For the hierarchical multi-condition fusion module, we implement six sparse-dense fusion blocks and two dense-dense fusion blocks, utilizing a base dilation step $s$ set to {4, 8, 12, 16, 20, 24}.
For diffusion process, we adopt the cosine schedule as beat schedule and set diffusion steps $t$ as 1000 diffusion following~\cite{mdm}. The classifier-free guidance scale is set to 2. The diffusion module aligns with a transformer-based architecture similar to~\cite{edge}. We adopt Adam optimizer with the learning rate set as 2e-4 and the training batch size is set to 64.
During training, we randomly sample between 1\% and 30\% of the ground truth (GT) motion frames as the keyframe conditions. This training strategy can effectively improve the generalization capability of our method on different numbers of keyframes. We use the GT motion beats as beat conditions.

To ensure fair comparisons, we follow EDGE~\cite{edge} to crop all data into 5-second clips at 30 frames per second, with 2.5 seconds of overlapping.
During testing, we randomly sample 10\% of keyframes from unpaired samples in the testing set as keyframe conditions and subsets of testing musical beats as beat conditions to prevent data leakage. The musical beats used for experiments are extracted by the off-the-shelf audio toolkit Librosa~\cite{librosa}.
All experiments are conducted on 4 NVIDIA RTX3090 GPUs.

\textbf{Evaluation Metrics.}
For quantitative evaluation, we measure the generated dance from three aspects: generation quality, diversity, and controllability. Due to the limited samples in the testing set of AIST++~\cite{aist++}, the prior work EDGE~\cite{edge} has demonstrated that the Fréchet Inception Distance (FID) is not reliable for measuring the generation quality. Therefore, we do not adopt FID as the evaluation metric.
Instead, we assess generation quality from two different perspectives: music-dance correlation and kinematic plausibility. In terms of music-dance correlation, we utilize the Beat Alignment Score (BAS) metric, following the approach presented in~\cite{bailando}. BAS quantifies the synchronization quality between the generated dance and the musical beat. For the evaluation of kinematic plausibility, we adopt the Physical Foot Contact score (PFC) metric proposed in~\cite{edge}. PFC provides a metric for measuring the physical plausibility of the generated dance from a kinematic perspective.
As for the diversity, we measure the average feature distance of kinetic ($\text{Div}_k$) and geometric ($\text{Div}_g$) features extracted by fairmotion~\cite{fairmotion} following the previous works~\cite{aist++, bailando, edge}.

For assessing controllability, we employ the Key Pose Distance (KPD) and Beat Assignment Precision (BAP) metrics to evaluate keyframe control and beat control, respectively.
KPD quantifies the average mean square error of local joint positions in Cartesian space at the keyframes.
To evaluate beat control, we measure the alignment between the generated dance and the designated beat condition. BAP is defined as the percentage of beats in the generated dance that are correctly assigned to the designated beat frames. Note that this metric differs from the BAS, as a dance sequence with a high BAS does not necessarily adhere to the specific motion beat choreography constraint we intend to impose.
\begin{table*}[t]
    \centering
    \caption{Quantitative comparisons among different methods on AIST++.
    \textbf{Bold} indicates best result. $\downarrow$ means lower is better, $\uparrow$ means higher is better and $\rightarrow$ means closer to the ground truth is better.}  
        \begin{tabular}{ccccccc}
        \toprule
                                    & \multicolumn{2}{c}{Quality} & \multicolumn{2}{c}{Diversity} & \multicolumn{2}{c}{Controllability} \\ \cmidrule(l){2-7} 
        \multirow{-2}{*}{Methods} & PFC $\downarrow$     & BAS $\uparrow$     & $\text{Div}_k \rightarrow$      & $\text{Div}_m \rightarrow$      & KPD $\downarrow$         & BAP $\uparrow$        \\ \midrule
        Ground Truth   & 1.338 & 0.384 & 9.773  & 7.212          & -     & - \\ \midrule
        FACT~\cite{aist++}           & 2.698 & 0.202 & 9.704  & 7.342          & -     & - \\
        Bailando~\cite{bailando}      & 1.578 & 0.215 & 9.622  & 7.175 & -     & - \\   \midrule
        EDGE~\cite{edge}(keyframes) & 1.084 & 0.235 & \textbf{9.743} & 7.274          & 0.859 & - \\
        \rowcolor{gray!20}Ours(beat \& keyframes)            & \textbf{0.966} & \textbf{0.661} & 9.660    & \textbf{7.248}      & \textbf{0.306}   & \textbf{0.793}   \\ \bottomrule
        \end{tabular}
    % }
    \label{comparison}
\end{table*}

\subsection{Comparison to Existing Methods}
Currently, only three music-to-dance methods are publicly accessible: EDGE~\cite{edge}, Bailando~\cite{bailando}, and FACT~\cite{aist++}. Of these, EDGE~\cite{edge} is most directly aligned with our approach, as it similarly accommodates keyframes as inputs for generating dance sequences. We therefore assess EDGE using the same keyframe conditions as our method to ensure a fair comparison. Conversely, certain approaches, like DanceFormer~\cite{danceformer}, are not open-source, thus hindering any direct evaluative comparison.
We use the original test split of AIST++~\cite{aist++} for evaluation, which has no overlap with the training split in terms of both the music and dance choreography.
\noindent\begin{minipage}[t]{.48\textwidth}
    \centering
    \captionof{table}{Results of user study.}
    \setlength{\tabcolsep}{0.05cm}{
        \resizebox{0.9\columnwidth}{!}{%
        \begin{tabular}{ccc}
        \toprule
        User Study & \multicolumn{2}{c}{Beat-It Win Rate} \\ \midrule
        FACT       & \multicolumn{2}{c}{92.2\%}        \\
        Bailando   & \multicolumn{2}{c}{78.8\%}        \\ \midrule
        \multirow{2}{*}{\makecell[c]{EDGE\\ (keyframes)}} & Quality & \begin{tabular}[c]{@{}c@{}}Controllability\\ (Keyframe)\end{tabular} \\ \cmidrule(l){2-3} 
                    & 60.3\%         & 86.9\%     \\ \bottomrule
        \end{tabular}
    }}
    \label{user_study}
\end{minipage}
\noindent\begin{minipage}[t]{.48\textwidth}
    \centering
    \captionof{table}{Ablation study on AIST++.}
    \resizebox{0.95\columnwidth}{!}{
    \renewcommand\arraystretch{1.1}
    \begin{tabular}{ccccc}
        \toprule
                                    & \multicolumn{2}{c}{Quality}       & \multicolumn{2}{c}{Controllability} \\ \cline{2-5} 
        \multirow{-2}{*}{Ablations} & PFC $\downarrow$ & BAS $\uparrow$ & KPD $\downarrow$  & BAP $\uparrow$  \\ \hline
        w/o HF                   & 25.626 & 0.322 & 0.477 & 0.323 \\
        w/o BD                   & 1.632 & 0.358  & 0.389 & 0.371 \\
        w/o $\mathcal{L}_{beat}$ & 1.342 & 0.397 & 0.343 & 0.411 \\
        \rowcolor{gray!20}
        Ours                     & \textbf{0.966} & \textbf{0.661} & \textbf{0.306} & \textbf{0.793} \\ \bottomrule
    \end{tabular}}
    \label{ablation}
\end{minipage}

\begin{figure}[t]
    \centering	
    \includegraphics[width=\linewidth,scale=1.00]{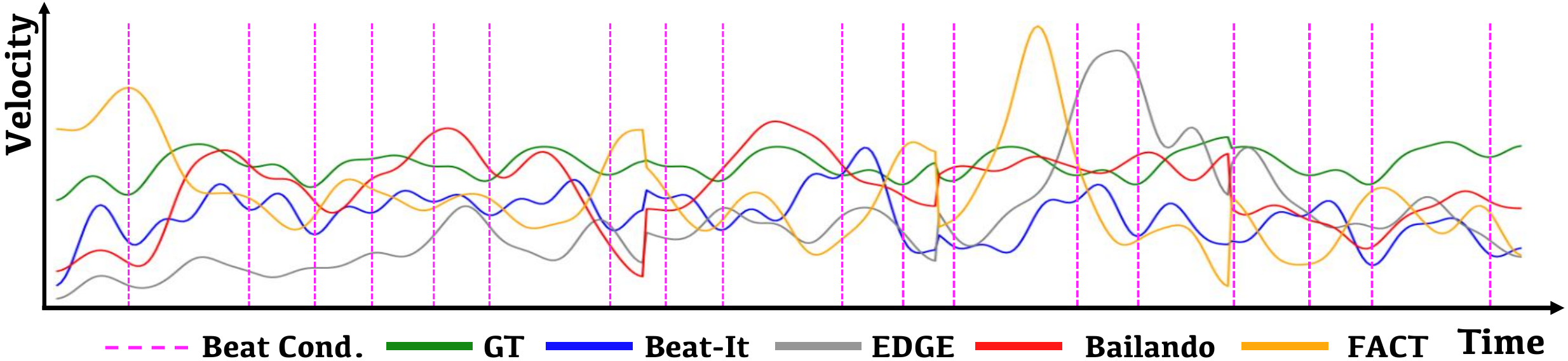}
    \caption{
        \textbf{Visualization comparison on beat alignment among different methods.} The motion generated by our method shows precise beat alignment with the given beat condition, demonstrating the superiority of our method in beat control.
    }	
    \label{beat_align_visualization}	
\end{figure}

\textbf{Quantitative Comparisons.} The quantitative results are shown in \reftab{comparison}. As is revealed, our method surpasses all the competitors in terms of generation quality, diversity, and controllability.
Particularly, our method shows considerable advantages over the compared methods for BAS metrics, notably with an improvement of 0.426 compared with the SOTA method EDGE~\cite{edge}. This demonstrates the effectiveness of our method in generating beat-synchronized dance sequences while maintaining diversity.
Furthermore, in the aspect of controllability, our method also significantly excels EDGE~\cite{edge} with a notable improvement of 0.553 in KPD. \reffig{beat_align_visualization} shows the mean joint velocities over time for a specific dance motion concerning both the GT and the existing methods.
Notably, compared with others, our approach demonstrates more motion beats (local minimum point of velocities) precisely at the beat frames. This contributes to a better visual alignment with the audio when compared to the other methods. The BAP further validates our method's superior ability to produce dance sequences precisely aligned with the beat condition. Due to limited space, we present additional quantitative experiments in the supplementary materials.

\textbf{User Study.} We also conduct a user study to assess the generation quality and controllability of our method. We randomly select 20 generated dance sequences from the testing set and ask 18 participants to pick the best one among our method and the other compared methods. In particular, participants are required to choose the superior dance according to several criteria, namely, the overall visual plausibility of movements, synchronization with the dance beats, and keyframe controllability. The results of the user study are tabulated in \reftab{user_study}. Our method shows noticeable advantages in terms of generation quality compared to other approaches, particularly in beat alignment. 
For qualitative evaluations on keyframe-conditioned controllability, we also provide the participants with the keyframe references for intuitive comparison between our method against EDGE~\cite{edge}. The results show that our approach owns better motion controllability than EDGE~\cite{edge}. 

\subsection{Ablation Studies}

To get a more comprehensive insight into our main contributions, we conduct ablation studies on the hierarchical multi-condition fusion module, the beat-aware dilation scheme, and the beat alignment loss to evaluate their efficacy. The quantitative results are shown in \reftab{ablation}. The qualitative results can be also found in the supplementary video.

\textbf{Hierarchical Multi-condition Fusion Module.}
To verify the significance of the hierarchical multi-condition fusion module, we remove it by simply concatenating all the conditions and feeding them into a one-stage encoder (w/o HF). As shown in \reftab{ablation}, the simple one-stage fusion module can lead to inferior results in both quality and controllability.
This is because the direct concatenation of multiple conditions can lead to inevitable condition conflicts, giving rise to suboptimal optimization of the model. Even worse, the naive padding of the sparse keyframe condition introduces severe noises, further complicating the learning process and resulting in unsatisfactory generation performance. By decently fusing multiple conditions gradually, our hierarchical multi-condition fusion mechanism effectively mitigates the conflicts among different conditions and yields more informative conditions, allowing for higher-quality generation with better controllability.

\textbf{Beat-aware Dilation Scheme.}
To validate the effectiveness of our beat-aware dilation scheme, we replace it with a vanilla masked attention strategy (w/o BD). As shown in \reftab{ablation}, the absence of the beat-aware dilation scheme notably impairs both keyframe and beat choreography controllability.
The reason behind this is that beat-aware dilation can dynamically adjust the expansion degree of the attention mask, contingent on the distance between the keyframe and the beat frames. This tactical approach significantly improves the synchronization between keyframes and their associated beat frames, thereby empowering the model to more effectively harness prior knowledge of beats. Consequently, this enhancement considerably bolsters beat control while augmenting keyframe controllability.

\textbf{Beat Alignment Loss.}
To further validate the effectiveness of the beat alignment loss, we assess the model's performance by comparing it to its counterpart without it (w/o $\mathcal{L}_{beat}$). As shown in \reftab{ablation}, the ablated variant without the beat alignment loss demonstrated performance degradation in beat choreography, showcasing decreases of 0.264 and 0.382 in BAS and BAP, respectively.
This decline in performance can be attributed to the advantageous impact of direct supervision during training, particularly in guiding the model to comprehend beat-level dance choreography through the proposed beat alignment loss. Consequently, this enhancement significantly improves the synchronization between motion sequences and designated beats.

\section{Discussion and Conclusion}
In this work, we propose a novel multi-condition diffusion-based framework, Beat-It, for beat-synchronized and key pose-guided dance generation.
The proposed framework presents a hierarchical multi-condition fusion mechanism equipped with a beat-aware mask dilation scheme to integrate conditions with different information sparsity. To achieve precise beat synchronization, we explicitly disentangle the beats from the music and inject beat controllability throughout the entire generation process. Additionally, we introduce a specifically designed beat alignment loss to provide explicit guidance and supervision on motion beats. 
Both qualitative and quantitative experimental results on the AIST++ dataset validate the superiority of our method in producing high-quality dance sequences with precise beat synchronization and flexible keyframe control.
Further limitations and broader impact are discussed in supplementary materials.

\section*{Acknowledgements}
The work is supported by China National Key R\&D Program (No. 2023YFE0202700), Key-Area Research and Development Program of Guangzhou City (No. 2023B01J0022), Guangdong Provincial Natural Science Foundation for Outstanding Youth Team Project (No. 2024B1515040010), Health and Medical Research Fund (HMRF) of Hong Kong Health Bureau (No. 10211516), Guangdong Natural Science Funds for Distinguished Young Scholar (No. 2023B1515020097), National Research Foundation Singapore under the AI Singapore Programme (No. AISG3-GV-2023-011). 

% ---- Bibliography ----
%
% BibTeX users should specify bibliography style 'splncs04'.
% References will then be sorted and formatted in the correct style.
%

\bibliographystyle{splncs04}
\bibliography{main}

% \end{sloppypar}
\end{document}

\sloppy
% ---------------------------------------------------------------
% TODO REVIEW: Replace with your title
\title{Beat-It: Beat-Synchronized Multi-Condition 3D Dance Generation\\\textemdash Supplementary Materials\textemdash}

% TODO REVIEW: If the paper title is too long for the running head, you can set
% an abbreviated paper title here. If not, comment out.
\titlerunning{Beat-It: Beat-Synchronized Multi-Condition 3D Dance Generation}

% % TODO FINAL: Replace with your author list. 
% Include the authors' OCRID for the camera-ready version, if at all possible.
\author{
    % Zikai Huang\inst{1}\orcidlink{0009-0005-4526-440X} \and
    % Xuemiao Xu\inst{1,3,4}$^{(\textrm{\Letter})}$\orcidlink{0000-0002-8006-3663}\and
    % Cheng Xu\inst{2}$^{(\textrm{\Letter})}$\orcidlink{0000-0002-4281-6214}\and
    % Huaidong Zhang\inst{1}\orcidlink{0000-0001-7662-9831} \and\\
    % Chenxi Zheng\inst{1}\orcidlink{0009-0006-0344-2439} \and
    % Jing Qin\inst{2}\orcidlink{0000-0002-7059-0929} \and
    % Shengfeng He\inst{5}\orcidlink{0000-0002-3802-4644}
}

% % TODO FINAL: Replace with an abbreviated list of authors.
\authorrunning{Z. Huang et al.}
% % First names are abbreviated in the running head.
% % If there are more than two authors, 'et al.' is used.

% % TODO FINAL: Replace with your institution list.
\institute{
% South China University of Technology, China \\
% \email{xuemx@scut.edu.cn}
% \and
% The Hong Kong Polytechnic University, Hong Kong SAR, China \\
% \email{cschengxu@gmail.com}
% \and
% Guangdong Engineering Center for Large Model and GenAI Technology \\
% \and
% Guangdong Provincial Key Lab of Computational Intelligence and Cyberspace Information\\
% \and
% Singapore Management University, Singapore\\
% \url{https://zikaihuangscut.github.io/Beat-It/}
}

\maketitle

%%%%%%%%% ABSTRACT

\section{Introduction} 
In this document, we first provide the details of the music and dance representations (\refsec{representation}). 
Next, we describe the architecture of the beat distance estimator (\refsec{estimator}). 
Then, additional quantitative experiments are provided for a comprehensive evaluation of our methods (\refsec{addition_exp}). 
Afterward, we show the qualitative results of our methods, including comparison results with existing approaches, ablation study, beat-synchronized keyframe-controlled dance generation, arbitrary beat-controlled dance generation, and the in-the-wild results (\refsec{results}). 
Finally, we discuss the limitations and broader impact of our work (\refsec{discussion}). 

\section{Representations of Music and Dance} 
\label{sec:representation}
In this section, we provide detailed explanations of the music and dance representations adopted in our method.

\textbf{Music Representation.}
Previous works show that using the music feature extracted from large pre-trained models as music condition brings significant improvements to music-to-dance generation~\cite{edge, diffdance}. Therefore, we adopt the same music feature representation as EDGE~\cite{edge}. This music feature is then passed through a transformer encoder $\mathcal{E}_m$ to produce the final music embedding $\mathbf{e}_m \in \mathbb{R}^{L\times d}$.

\textbf{Dance Representation.}
Similar to prior works~\cite{mnet, danceformer, aist++, edge, neverstop, diffdance, gdancer}, we adopt the rotation matrix to represent dance motions. In particular, we use the root translation $\delta$ and 6-DOF rotation representation $r$~\cite{6dof} for 24 joints in SMPL format~\cite{smpl} for each frame. Following EDGE~\cite{edge}, we also include a binary contact label $g$ for the heel and toe of each foot. The entire pose representation is therefore denoted as $\mathbf{x}^i = \{g, \delta, r\} \in \mathbb{R}^{D=4+3+144=151}$. 
Correspondingly, the keyframe condition embedding $\mathcal{X}^{ref} \in \mathbb{R}^{L \times 151}$ can be projected into a keyframe embedding $\mathbf{e}_r \in \mathbb{R}^{L \times d}$ via the keyframe encoder $\mathcal{E}_r$. To relieve the negative impacts caused by the sparse nature of the keyframes, we explicitly introduce positional context information by adding learnable embeddings of the nearest keyframe distance to $\mathbf{e}_r$ at the non-keyframe locations.

\section{The Beat Distance Estimator}
\label{sec:estimator}
The beat distance estimator is used to estimate the nearest beat distance of each frame for loss calculation. It is pre-trained on a beat distance regression task. In particular, it comprises a 6-layer 4-head transformer-based encoder with hidden size of 128.
Given a sequence of motion as input, it is first processed through forward kinematics to obtain the 3D joint positions. Then we get the joint velocity and feed it into the beat distance estimator, followed by an MLP, to predict the nearest beat distance of each frame. 

\section{Additional Quantitative Experiments}
\label{sec:addition_exp}
Here we provide quantitative comparison between our method and the state-of-the-art motion interpolation approach, and the performance of our method with different keyframe condition ratios, different single conditions and different condition combinations.

\textbf{Comparison with Motion Interpolation Method.} For reference, we also present the comparison between our method and the state-of-the-art motion interpolation method~\cite{mib} quantitatively. But note that the motion-interpolation task accepts key poses as the only input condition, which is essentially different from our setting of dance generation dominated by music. The results in Tab.~\ref{suppl_mib} indicate that it is not feasible to directly apply the motion interpolation method to render beat-synchronized dance motions due to its lack of music and beat guidance. In comparison, our method demonstrates superior capability of generating high-quality beat-synchronized dance motions.
\begin{table}[h]
    \caption{Quantitative comparison with Qin et al.~\cite{mib} on AIST++~\cite{aist++}.}
    \centering
    \begin{tabular}{cccccc}
        \toprule
        & \multicolumn{2}{c}{Quality} & \multicolumn{2}{c}{Diversity} & \multicolumn{1}{c}{Controllability} \\ \cmidrule(l){2-6} 
        \multirow{-2}{*}{Methods}     & PFC $\downarrow$     & BAS $\uparrow$     & $\text{Div}_k \rightarrow$      & $\text{Div}_m \rightarrow$     & BAP $\uparrow$        \\ \midrule
        Ground Truth                  & 1.338                & 0.384              & 9.773                          & 7.212                           & - \\ \midrule
        Qin et al.~\cite{mib}         & 3.790                & 0.182              & 9.7981                         & 7.112                           & - \\ 
        Ours                          & \textbf{0.966}       & \textbf{0.661}     & \textbf{9.660}                 & \textbf{7.248}                  & \textbf{0.793}   \\ \bottomrule
    \end{tabular}
    \label{suppl_mib}
\end{table}

\textbf{Different Ratios of Keyframes Condition.} Additionally, we also report the performance of our method with varying ratios of keyframe conditions. The results, presented in \reftab{suppl_kf}, reveal that our approach can still produce compelling outcomes even if the input keyframes are highly sparse (e.g., $5\%$ or no keyframes). 
\begin{table}[h]
\caption{Quantitative results with different ratios of input keyframes on AIST++\cite{aist++}.}
    \centering
        \begin{tabular}{ccccccc}
        \toprule
                                    & \multicolumn{2}{c}{Quality} & \multicolumn{2}{c}{Diversity} & \multicolumn{2}{c}{Controllability} \\ \cmidrule(l){2-7} 
        \multirow{-2}{*}{\makecell[c]{Keyframes \\Ratio}} & PFC $\downarrow$     & BAS $\uparrow$     & $\text{Div}_k \rightarrow$      & $\text{Div}_m \rightarrow$      & KPD $\downarrow$         & BAP $\uparrow$        \\ \midrule
        Ground Truth   & 1.338  & 0.384  & 9.773   & 7.212   & -      & - \\ \midrule
        0\%            & 1.157  & 0.644  & 11.298  & 7.310   & -      & 0.782   \\
        5\%            & 0.758  & 0.738  & 9.510   & 7.123   & 0.301  & 0.794   \\ 
        10\%           & 0.966  & 0.661  & 9.660   & 7.248   & 0.306  & 0.793   \\ \bottomrule
        \end{tabular}
    \label{suppl_kf}
\end{table}

\textbf{Different Single Conditions.} Our method supports single conditional generation by directly setting the other conditions to ``null condition” during the training process.
\reftab{suppl_single_condition} presents the quantitative results of our model with different single conditions. We can observe that our music-only model has already shown significant advantages over the SOTAs. By integrating all three conditions, the overall performance of our full model is further enhanced, demonstrating considerable advantages over the variants with single input condition. Importantly, models with beat- or keyframe-only condition violate the original music-to-dance generation setting and generate dance sequences with severe freezing issue, thus leading to significantly lower PFC values.
\begin{table}[h]
    \caption{Quantitative comparison with different single conditions.}
    \centering
    \resizebox{0.95\linewidth}{!}{
    \begin{tabular}{ccccccc}
        \toprule
        & \multicolumn{2}{c}{Quality} & \multicolumn{2}{c}{Diversity} & \multicolumn{2}{c}{Controllability} \\ \cmidrule(l){2-7}
        \multirow{-2}{*}{Methods}    & PFC $\downarrow$     & BAS $\uparrow$     & $\text{Div}_k \rightarrow$      & $\text{Div}_m \rightarrow$  & KPD $\downarrow$         & BAP $\uparrow$        \\ \midrule
        Ground Truth                 & 1.338                & 0.384              & 9.773                           & 7.212                       & -                        & -                     \\ \midrule
        FACT~\cite{aist++}          & 2.698                & 0.202              & 9.704                           & 7.342                       & -                        & -                     \\
        Bailando~\cite{bailando}    & 1.578                & 0.215              & 9.622                           & 7.175                       & -                        & -                     \\
        EDGE (keyframes)~\cite{edge} & 1.084                & 0.235              & 9.743                           & 7.274                       & 0.859                    & -                     \\ \midrule
        Ours (beats only)            & \textbf{0.058}       & 0.557              & 10.078                          & 4.383                       & -                        & 0.586                 \\
        Ours (keyframes only)        & 0.443                & 0.203              & 11.124                          & 6.783                       & 0.673                    & -                     \\
        Ours (music only)            & 0.879                & 0.237              & \textbf{9.782}                           & 6.997                       & -                        & -                     \\
        \rowcolor{gray!20}Ours (music $+$ beat $+$ keyframes)            & 0.966     & \textbf{0.661}    & 9.660    & \textbf{7.248}      & \textbf{0.306}   & \textbf{0.793}   \\
        \bottomrule
    \end{tabular}}
    \label{suppl_single_condition}
\end{table}

\textbf{Different Conditions Combinations.} Our framework can be easily extended to support different conditions combinations by directly setting the omitted condition to ``null condition'' during the training process. Quantitative results of our methods with various condition combinations are tabulated in \reftab{suppl_conds}. The results affirm the versatility and flexibility of our approach in supporting diverse condition combinations. 
\begin{table}[h]
    \caption{Quantitative results under different combinations of conditions.}
    \centering
    \begin{tabular}{ccccccc}
        \toprule
        & \multicolumn{2}{c}{Quality} & \multicolumn{2}{c}{Diversity} & \multicolumn{2}{c}{Controllability} \\ \cmidrule(l){2-7} 
        \multirow{-2}{*}{Methods}   & PFC $\downarrow$     & BAS $\uparrow$     & $\text{Div}_k \rightarrow$      & $\text{Div}_m \rightarrow$      & KPD $\downarrow$         & BAP $\uparrow$        \\ \midrule
        Ground Truth                          & 1.338       & 0.384       & 9.773       & 7.212       & -          & - \\ \midrule
        music $+$ keyframes                   & 0.680       & 0.240       & 9.487       & 7.145       & 0.304      & - \\
        music $+$ beats                       & 1.157       & 0.644       & 11.298      & 7.310       & -          & 0.782 \\ 
        music $+$ keyframes $+$ beats (Ours)  & 0.966       & 0.661       & 9.660       & 7.248       &0.306       & 0.793   \\ 
        \bottomrule
    \end{tabular}
    \label{suppl_conds}
\end{table}

\section{Qualitative Results}
\label{sec:results}
For a comprehensive evaluation of our method, we include all the qualitative results in the supplementary demo video and make them accessible on \textit{\href{https://zikaihuangscut.github.io/Beat-It/}{our website}} for ease of reference. We recommend viewing these materials with audio enabled for the best experience.

\textbf{Comparison with Existing Methods.}
We compare our method with the state-of-the-art methods EDGE~\cite{edge}, Bailando~\cite{bailando}, and FACT~\cite{aist++} on the AIST++~\cite{aist++} dataset.
Note that EDGE~\cite{edge} is the one and only open-source keyframe-conditioned dance generation method.
Bailando~\cite{bailando} utilizes 3D Cartesian space key points for dance representation, which are not directly applicable for rendering 3D characters. To ensure a fair comparison, we directly utilize skeleton stick figures for visualization. The results indicate a remarkable improvement in both the quality and controllability of our method compared to these existing techniques.
Please refer to the supplementary demo video for the full comparison results. 

\textbf{Ablation Study.}
As shown in the qualitative results of the ablation study in the supplementary demo video. 
Once the hierarchical multi-condition fusion module is removed (w/o HF), a notable degradation in performance with less coherent generated dance movements can be observed. This is mainly attributed to the direct concatenation of multiple conditions, which introduces conflicts among conditions and hampers optimal model optimization. In contrast, our proposed hierarchical multi-condition fusion mechanism effectively alleviates conflicts between conditions and generates more informative conditions for better guidance.
Additionally, the absence of the beat-aware mask dilation scheme (w/o BD) causes significant jittering in the generated motions and significantly impairs keyframe controllability and beat synchronization. The rationale behind this is the inherent dynamic adjustment capability of beat-aware dilation for the adaptive expansion of attention masks, which is crucial for enhancing synchronization between keyframes and their associated beat frames, allowing the model to better leverage prior knowledge of beats. 
Omitting the beat alignment loss (w/o $\mathcal{L}_{beat}$) results in generation with inferior beat alignment, demonstrating the significance of the explicit supervision signals from beats on improving beat synchronization. 

\textbf{Beat-Synchronized Keyframe-Controlled Dance Generation.}
To further demonstrate the superiority of our method in beat-synchronized keyframe-controlled dance generation, we conducted experiments using identical music and beats while varying the keyframes. The results, showcased in the supplementary demo video, highlight the remarkable ability of our model to generate dance sequences that not only adhere to specified keyframes but also precisely align with given beat conditions. To visually depict beat synchronization, mean joint velocity curves of the generated dance motions are incorporated, aligning with the provided beats indicated by vertical dashed lines. A noteworthy feature of our method is its proficiency in producing motion beats, identified as local minimum velocity points, precisely aligned with the beat condition. This alignment serves as a clear indication of our method's advanced capability in achieving beat synchronization. 

\begin{figure}[h]
    \centering	
    \includegraphics[width=.9\linewidth]{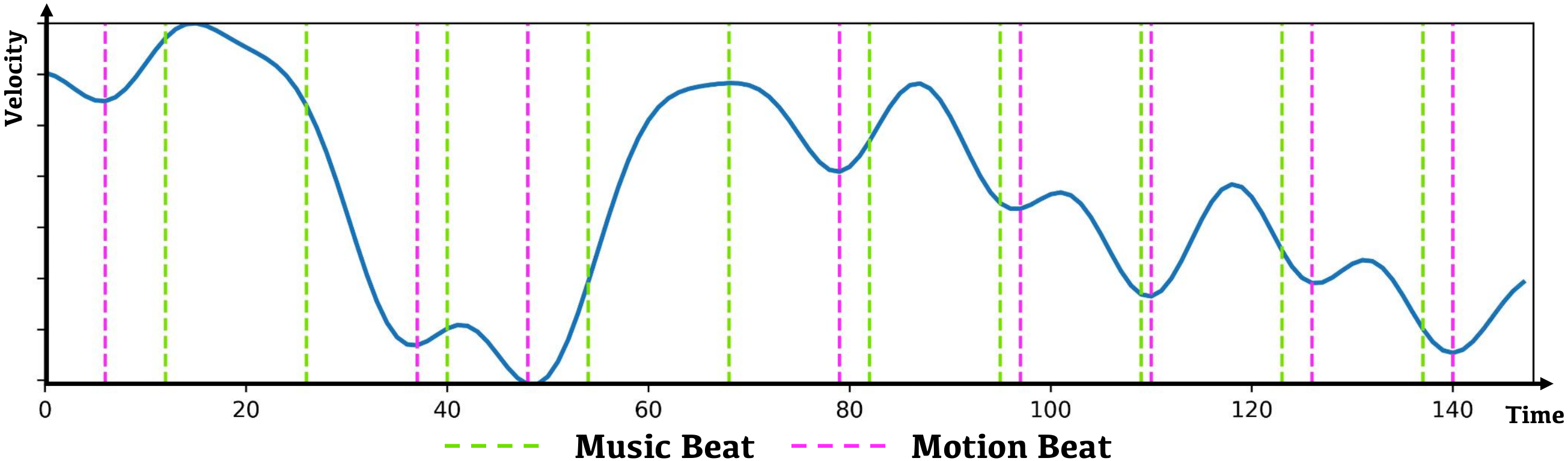}
    \caption{Visualization of motion beat alignment given a beat condition not strictly aligned with the musical beats. }	
    \label{ballet_beat_motion}	
\end{figure}

\textbf{Arbitrary Beat-Controlled Dance Generation.} 
In practical choreography, the arrangement of dance motion beats is highly diverse. Apart from the straightforward incorporation of a subset of music beats as motion beats, dance generation without strict alignment with the musical beats is also needed in some special circumstances (e.g., the synchrony between the musical beats and dance motions is often less pronounced in ballet.). Thanks to the explicit disentanglement of beat conditions, our approach enables dance generation with diverse motion beats. These beats can be controlled by any specified beat conditions, offering flexibility beyond strict adherence to the musical beats. To demonstrate this, we illustrate an example where the beat condition is not strictly aligned with the music beats. The generated dance motions, depicted in \reffig{ballet_beat_motion}, confirm the effectiveness of our method in creating convincing results based on the specified motion beat condition, rather than by strict synchronization with the music beats. This underscores the significant generalizability of our approach to various dance genres with diverse beat patterns. 

\textbf{In-the-Wild Results.}
To further demonstrate the generalization ability of our method, we also evaluate our method on in-the-wild music videos. Specifically, we randomly select different samples from the AIOZ-GDANCE~\cite{aiozGdance} dataset for evaluation. This dataset exhibits much higher diversity than our training data of AIST++~\cite{aist++}. 
Our method can generate high-quality dance motions that are well-aligned with the given beats and keyframes.

\section{Limitation and Broader Impact}
\label{sec:discussion}
\hspace{\parindent}\textbf{Limitation.} Similar to all diffusion-based methods (\eg, EDGE~\cite{edge}), a general limitation of our method is the inference speed. 
Here we show the model size and runtime comparison of different methods for generating a 150-frame dance sequence in \reftab{rebuttal_complexity}. Although our method involves increased computation time due to our multi-modal fusion and diffusion process, it establishes the first highly controllable dance generation paradigm, significantly surpassing existing methods in terms of generation quality, flexibility, and controllability, which are crucial for practical application. We believe our extra overhead is acceptable and can be further reduced via a more advanced sampling strategy or model distillation.
While our method can be extended to dances with conditions in the wild, the sliding foot issue may arise in some extremely complex scenarios. The integration of a more sophisticated loss function based on physical kinematics holds promise for further enhancing the overall quality of generated dance sequences.
\begin{table}[h]
    \centering
    \caption{Complexity comparison of different methods.}
    \begin{tabular}{ccccc}
        \toprule
        Metrics         & FACT~\cite{aist++}     & Bailando~\cite{bailando}     & EDGE~\cite{edge}      & Ours     \\ \midrule
        Params/M        & 120.4         & 173.4             & 49.5           & 106.4    \\
        Runtime/s       & 22.4          & 0.3               & 1.6            & 6.0      \\
        \bottomrule
    \end{tabular}
    \label{rebuttal_complexity}
\end{table}

\textbf{Broader Impact.} Our research has potential positive impacts on the dance community, providing choreographers with a new tool for fine-grained dance creation, thereby supporting artistic endeavors. On the other hand, it may also lead to adverse societal impacts. Automated dance generation poses risks of copyright infringement, cultural appropriation, and devaluation of originality in dance creation. Addressing these issues requires ethical development within the dance community, emphasizing cultural sensitivity, and maintaining appreciation for human creativity.

%%%%%%%%% REFERENCES
\bibliographystyle{splncs04}
\bibliography{main}